\documentclass[11pt,preprint]{aastex}

\def\lya{\ifmmode {{\rm Ly}\alpha}\else
	Lyman-$\alpha$\fi}

\def\erg{\hbox{erg}}

\def\f17{f_{17}}
\def\lsun{L_\odot}
\def\Msun{M_\odot}
\def\Mpc{\hbox{Mpc}}
\def\nm{\hbox{nm}}

\def\kms{\hbox{\,km s$^{-1}$}}
\def\year{\hbox{yr}}
\def\Myr{\hbox{Myr}}

\def\arcsec{\ifmmode {''}\else{$''$}\fi}
\def\microJy{\mu\hbox{Jy}}
\def\zre{z_r}
\def\fesc{f_{\rm esc}}

\def\ergcm2s{\ifmmode {\rm\,erg\,cm^{-2}\,s^{-1}}\else
                ${\rm\,ergs\,cm^{-2}\,s^{-1}}$\fi}
\def\ergsec{\ifmmode {\rm\,erg\,s^{-1}}\else
                ${\rm\,ergs\,s^{-1}}$\fi}
\def\kmsMpc{\ifmmode {\rm\,km\,s^{-1}\,Mpc^{-1}}\else
                ${\rm\,km\,s^{-1}\,Mpc^{-1}}$\fi}

\def\oiipair{[\ion{O}{2}] $\lambda \lambda$3726,3729}

\def\oiiipair{[\ion{O}{3}] $\lambda \lambda$4959,5007}

\begin{document}

\title{
Spectroscopic Confirmation of Three Redshift $z \approx 5.7$ Lyman-$\alpha$
Emitters from the Large Area Lyman Alpha Survey
}

\author{
James E. Rhoads\altaffilmark{1,2},
Arjun Dey\altaffilmark{3},
Sangeeta Malhotra\altaffilmark{1,2},
Daniel Stern\altaffilmark{4},
Hyron Spinrad\altaffilmark{5},
Buell T. Jannuzi\altaffilmark{3},
Steve Dawson\altaffilmark{5},
Michael Brown\altaffilmark{3}
}

\begin{abstract}
Narrow-band searches for \lya\ emission are an efficient way of
identifying star-forming galaxies at high redshifts.  We present
Keck telescope spectra confirming redshifts $z\approx 5.7$ for three
objects discovered in the Large Area Lyman Alpha (LALA) survey at Kitt Peak
National Observatory.
All three spectra show strong, narrow emission lines with the
asymmetric profile that is characteristically produced in high
redshift \lya\ emitters by preferential HI absorption in the blue wing
of the line.  These objects are undetected in deep $B_W$, $V$, $R$,
and $\lambda \approx 6600$\AA\ narrow-band images from the NOAO Deep
Wide-Field Survey and from LALA, as expected from Lyman break and
\lya\ forest absorption at redshift $z\approx 5.7$.  All three objects
show large equivalent widths ($\ga 150$\AA\ in the rest-frame),
suggesting at least one of the following: a top-heavy initial mass
function, very low stellar metallicity, or the presence of an active nucleus.
We consider the case for an active nucleus to be weak
in all three objects due to the limited width of the \lya\ 
emission line ($< 500 \kms$) and the absence of any other
indicator of quasar activity.
The three confirmed high redshift objects were among four spectroscopically
observed targets drawn from the sample of 18 candidates presented by Rhoads
\& Malhotra (2001).  Thus, these spectra support the \lya\
emitter population statistics from our earlier photometric study,
which imply little evolution in number density from $z=5.7$ to $z=4.5$ and
provide strong evidence that the reionization redshift $\zre > 5.7$.
\end{abstract}

\altaffiltext{1}{Space Telescope Science Institute, 3700 San
 Martin Drive, Baltimore, MD 21218. Email: rhoads@stsci.edu, san@stsci.edu} 
\altaffiltext{2}{Visiting Astronomer, Kitt Peak National Observatory,
National Optical Astronomy Observatory, which is operated by the
Association of Universities for Research in Astronomy, Inc. (AURA)
under cooperative agreement with the National Science Foundation.}
\altaffiltext{3}{Kitt Peak National Observatory, National Optical Astronomy
Observatory, which is operated by the Association of Universities
for Research in Astronomy, Inc. (AURA) under cooperative agreement
with the National Science Foundation.  Email: dey@noao.edu, jannuzi@noao.edu,
mbrown@noao.edu}
\altaffiltext{4}{Jet Propulsion Laboratory, California Institute of 
Technology, Mail Stop 169-327, Pasadena, CA 91109.
Email: stern@zwolfkinder.jpl.nasa.gov}
\altaffiltext{5}{
Department of Astronomy, University of California at Berkeley, Mail Code
3411, Berkeley, CA 94720 USA.  Email: spinrad@astro.berkeley.edu,
sdawson@astro.berkeley.edu}

\section{Introduction}

Narrow-band searches were originally proposed as a method of finding
young, high redshift galaxies in their first throes of star-formation (Partridge \&
Peebles 1967). This method has now proven its worth in discovering
star-forming galaxies at ever increasing redshifts. A large fraction
(9 out of 11) of known $z > 5$ galaxies have strong \lya\ emission
(Dey et al. 1998, Dawson et al. 2002, Weyman et al. 1998, Hu et
al. 1999, Hu et al. 2002, this paper), the only exception being a pair
of Lyman break galaxies at z=5.34 (Spinrad et al 1998).  This is partly
due to the ease of confirming galaxies with strong emission lines and
partly because one can avoid night sky lines by using narrow-band
selection methods.  Ultimately, the \lya\ emitters at high redshifts
will provide useful constraints on models of galaxy formation and
evolution. A more immediate application is to constrain the epoch of
reionization. In a neutral universe, \lya\ photons from galaxies are
resonantly scattered into large, diffuse halos, whose extremely low
surface brightness will make their detection very challenging.  Thus,
the transition to a neutral universe will be marked by a sharp drop
in \lya\ galaxy counts.

In the Large Area Lyman Alpha (LALA) survey (Rhoads et al. 2000, Paper
I) we use narrow-band imaging with the wide-field Mosaic cameras at
Kitt Peak National Observatory (KPNO) and Cerro Tololo Inter-American
Observatory (CTIO) 4-meter telescopes to image a relatively large area
($36'\times 36'$) of the sky at once. This allows us to obtain
good statistics on the candidates selected by imaging, and 
use them to study galaxy formation and evolution at high
redshifts. Confirming candidates from narrow-band imaging by
spectroscopy requires large telescopes and is a slower process.  But
as we show in this and companion papers, the success rate for
spectroscopic confirmations is high, so conclusions drawn from mere
imaging candidates are probably accurate (Rhoads \& Malhotra 2001,
hereafter RM01; Malhotra \& Rhoads 2002; Ouchi et al. 2002).

\section{Observations}
\subsection{Narrow-band and Broad-band Imaging}
The LALA survey concentrates on two primary fields in constellations
Bo\"{o}tes (J142557+3532) and Cetus (J020520-0455).  Each field is $36
\times 36$ square-arcminutes in size, corresponding to a single pointing
of the 8192x8192 pixel NOAO Mosaic CCD cameras that we have used for
all of our imaging.  The LALA fields are located within the NOAO Deep
Wide-Field Survey (Jannuzi \& Dey 1999), which provides deep broad-band
images in $B_W$, $R$, and $I$ filters that support our narrow-band data.

The $z \approx 5.7$ search has been conducted in the Bo\"{o}tes field
using two narrow-band filters with central wavelengths $\lambda_c$ of
815 and 823~nm, full widths at half maximum (FWHM) transmission
$7.5\nm$, and peak throughputs $\approx 90\%$.  The wavelengths were
chosen to fall in a gap in the night sky emission line spectrum, thus
minimizing sky noise (Figure 1).  The filter bandpasses deteriorated
below these specifications outside a central circle of $\sim 30'$
diameter, resulting in an effective search area of $\approx 710$
square arcminutes.  Within this area, we identified a sample of 18
candidate \lya\ emitters based on strong ($> 5 \sigma$) detections in
a narrow-band filter that substantially exceed the continuum flux
level (observer-frame equivalent width $>80$\AA\ and a $4\sigma$
excess over $I$ band), and non-detections ($<2\sigma$) in NDWFS $B_W$
and LALA $V$ filter images.  This sample is presented in RM01, to which
we refer the reader for more details on the photometry and selection
criteria.

\subsection{Spectroscopic Observations}
Spectroscopic followup of four LALA $z=5.7$ candidates was conducted
at the Keck~I 10m telescope
on UT 2002 March 12, 13, and 14. The observations were made using the
Low Resolution Imaging Spectrometer (LRIS; Oke et al. 1995) equipped
with the 400 lines/mm grating ($\lambda_b=8500$~\AA) and multislit
masks with 1\farcs4 wide slitlets. The 4 LALA z=5.7 candidates were
targeted using 2 multislit masks; LALA5 1-05 and 1-06 were observed
on March 12 for $\approx$ 110 minutes, and LALA5 1-01 and 1-03 were
observed on March 13 and 14 for a total of 175 minutes. All
observations were made with the slitlets oriented at roughly
320$^\circ$ (east of north).  Seeing on the nights varied between
0\farcs7 to 1\farcs0.
Because the slits were wider than the seeing FWHM, spectral resolution
becomes a function of source size, from $\sim 6.5$\AA\ for unresolved
sources in the best seeing data to $12$\AA\ for sources that fill
the slit.  The LALA targets on these masks are all $\la 1''$ in size
(based on the narrow-band detection images) implying an effective
resolution in the range $6$ to $9$\AA.
 
These data were reduced using a combination of standard IRAF ONEDSPEC and
TWODSPEC tasks together with homegrown slitmask reduction tasks developed
by  Stern, Bunker, \& Stanford (private communication) for reducing
LRIS slit mask data.\footnote{Available at
{\tt http://zwolfkinder.jpl.nasa.gov/$\sim$stern/bogus.html}.}

\section{Results}
Three of the four $z\approx 5.7$ candidates were confirmed as \lya\
emitters with the expected redshifts and line fluxes.  We designate
these as LALA~J142546.76+352036.3, LALA~J142647.16+353612.6, and
LALA~J142630.34+354022.5.  These correspond respectively to candidate IDs
LALA5~1-03, LALA5~1-06, and LALA5~1-05 from RM01. The extracted
spectra of the candidates are presented in Figure~1. The photometric
and spectroscopic properties of these objects are summarized in
tables~\ref{phottab} and~\ref{spectab}.

The fourth candidate on the slit masks (LALA5 1-01) had the lowest
narrow-band flux among the four, corresponding to a line flux of
$(1.57 \pm 0.27)\times 10^{-17} \ergcm2s$.  This candidate was not
detected in the spectra to a $5\sigma$ upper limit of $\sim 0.7 \times
10^{-17} \ergcm2s$.  Since it was a single-filter detection in the
photometric survey, it was most likely a noise spike in the
narrow-band image and not a real object.  Given the large number
($\sim 10^{6.5}$) of independent resolution elements in the images, we
expect about 1 false positive at the $5\sigma$ level per LALA
imaging field per narrow-band
filter, regardless of the number of true emission line galaxies
present in the survey.  The number of false positives could be larger
if the noise properties of the image are not precisely gaussian.
A transient origin for this source is unlikely, as it is not
detected in the z' filter image that was obtained during the
same 4 night observing run as the 815 and 823~nm images.

\begin{deluxetable}{lllll}
\tablecolumns{5}
\tablewidth{0pc}
\tablecaption{Photometry}
\tablehead{
\colhead{ID} & \colhead{Photometric}
 & \colhead{W$_\lambda^{\rm rest} (2\sigma)$} & \colhead{I flux}
 & \colhead{z' flux}\\
 & \colhead{line flux\tablenotemark{a}}
 & \colhead{limit (\AA)} & \colhead{($\microJy$)} & \colhead{($\microJy$)} }
\startdata
LALA J142546.76+352036.3 & $1.82 \pm 0.27$ &
 $>1650$ &$ -0.07 \pm 0.05$ & $-0.10 \pm 0.11$\\
LALA J142647.16+353612.6 & $3.67 \pm 0.31$ &
 $>195$ & \phs$0.01 \pm 0.05$ & \phs$0.34 \pm 0.11$ \\
LALA J142630.34+354022.5 & $2.32 \pm 0.32$ &
 $>150$ &$ -0.01 \pm 0.04$ & \phs$0.16 \pm 0.11$ \\
\enddata
\tablenotetext{a}{Units of $10^{-17} \ergcm2s$.}
\tablecomments{Photometric properties of the three confirmed
$z\approx 5.7$ LALA
sources. The $2\sigma$ lower bounds on equivalent width are derived
from the narrow- and broad-band photometry and corrected to the
spectroscopically determined redshift. They are not corrected for IGM
absorption, which will affect both the broad and narrow-band
photometry approximately equally if the unabsorbed line profiles are
centered on zero velocity. \label{phottab} }
\end{deluxetable}

\begin{deluxetable}{lllllll}
\tablecolumns{7}
\tablewidth{0pc}
\tablecaption{Spectroscopy}
\tablehead{
\colhead{ID} & \colhead{Redshift} & \colhead{Spectroscopic} & 
\colhead{FWHM} & \colhead{$\Delta v$} & 
\multicolumn{2}{c}{Continuum ($\microJy$)\tablenotemark{b}}\\
 & & \colhead{line flux\tablenotemark{a}} & (\AA)
 & \colhead{(km/s)} & \colhead{Blue side} & \colhead{Red side}}
\startdata
LALA J142546.76+352036.3 & 5.746 &
$1.87 \pm 0.14$ & 11.7 & $\la 360$ & $0.010 \pm 0.011$ & $0.058\pm 0.028$ \\
LALA J142647.16+353612.6 & 5.700 &
 $3.88 \pm 0.15$ & 14 & $\la 460$ & $0.013 \pm 0.012$ & $0.115\pm 0.029$ \\
LALA J142630.34+354022.5 & 5.674 &
 $2.65 \pm 0.14$ & 8.8 & $\la 220$ & $0.008 \pm 0.012$ & $0.018\pm 0.028$ \\
\enddata
\tablenotetext{a}{Units of $10^{-17} \ergcm2s$.}
\tablenotetext{b}{Quoted uncertainties only account for random
(photon counting) errors and exclude possible systematic errors.
Red and blue side measurements are taken in $1200$\AA\ wide windows
beginning $\approx 30$\AA\ from measured line center.}
\tablecomments{Spectroscopic properties of the three confirmed
$z\approx 5.7$ LALA sources.  The redshift is based on the peak of the
observed line profile, which may slightly overestimate the redshift
since the blue wing of Lyman-$\alpha$ is absorbed.  The measurement
error on the line peak is estimated at $\delta z \approx \pm 0.002$.
The full width at half maximum (FWHM) is directly
measured from the line profiles.  The corresponding velocity width
$\Delta v$ is based on subtracting in quadrature the
effective instrumental resolution for a point source, and is therefore
an upper limit since the galaxies may have angular size comparable to
the $0.7''$ seeing prevalent in these data.  The measured line fluxes
agree very well with the photometrically inferred fluxes in
table~\ref{phottab}, except for LALA J142630.34+354022.5, whose line flux 
falls partially outside the narrow-band filter and is therefore
underestimated from the narrow-band data.
\label{spectab} }
\end{deluxetable}

All of the confirmed sources show asymmetric spectral lines, with a
sharp drop on the blue side and a more extended wing on the red side.
This asymmetry is characteristic of high redshift \lya\ and is
physically associated with absorption by neutral hydrogen in outflows
from actively star-forming galaxies (e.g., Stern \& Spinrad 1999;
Manning et al 2000; Dawson et al 2002).  This asymmetry is good
evidence against an \oiipair\ line identification, because the close
doublet structure of \oiipair\ is expected to give the opposite
asymmetry from \lya\ under normal physical conditions for [\ion{O}{2}].	
None of the sources show any clearly detected spectral line besides
the primary $\lambda \approx 8200$\AA\ line.  (The coverage is $6850
{\rm \AA} < \lambda < 10640$\AA\ for LALA5~1-05 and 1-06, and
$5584 {\rm \AA} < \lambda < 9400$ \AA\
for LALA5~1-03.)  This immediately rules out the \oiiipair\ doublet as a
possible contaminant.

Because the slit size exceeds the apparent angular size of the
sources, the measured line widths contain contributions from the
physical line width, the angular size of the source, and the seeing.
An upper limit to the physical line width can be obtained by assuming
the objects are unresolved at the 0\farcs7 seeing of the spectroscopic
observations, providing an effective instrumental resolution of $\sim
6$\AA.
The resulting line widths are given in Table~\ref{spectab}.
All are $< 500 \kms$.

None of the three sources shows detectable continuum on the blue side
of the line in the spectra.  There are marginal detections of the
continuum emission on the red side of the line in two of the three
sources in our spectra, with formal significance levels of $2 \sigma$
and $4\sigma$.  Thus, in two of the three sources we have evidence for
a continuum break across the line, again supporting the identification
as \lya. These continuum measurements were done using a
variance-weighted average of the flux in $1200$\AA\ wide windows on
either side of the lines.  These quoted significance levels are based
only on random (photon counting) error, and should be regarded as
upper bounds on the true significance, since systematic errors 
will increase the uncertainty in the continuum flux.

\begin{figure}[ht]
\plotone{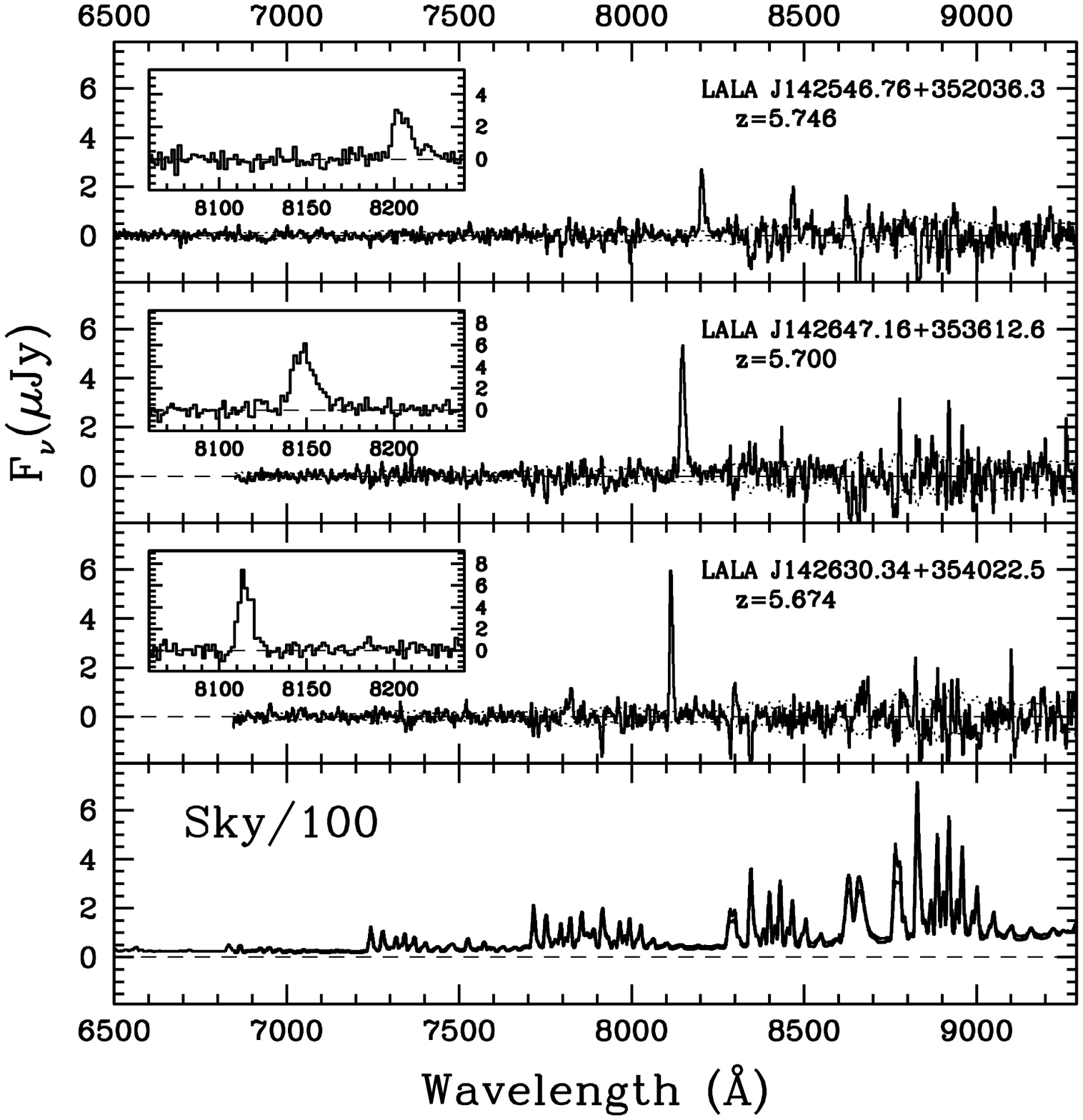}
\caption{ Spectra of the three confirmed $z\approx 5.7$ \lya\
emitters.  The top three panels show spectra smoothed with a Gaussian
of 4\AA\ FWHM (comparable to the spectral resolution).  The photon
counting errors (accounting for the applied smoothing) are also
plotted.  Insets show the primary emission line, and are unsmoothed to
allow line asymmetry to be studied.  Finally, the fourth panel shows
the night sky spectrum, reduced by a factor of 100 in order to plot it
on the same scale as the data.  (Night sky spectra for all three
objects are in fact overplotted here; the differences among them are
slight.)
\label{onedfig}}
\end{figure}

\section{Discussion} \label{disc}
We now consider the physical implications of our observations,
beginning with the properties of the individual sources and then
moving on to properties of the galaxy population and of the universe
at $z\approx 5.7$.  

The three spectroscopically confirmed \lya\ sources presented here all
show very large equivalent widths, with $2\sigma$ lower bounds of
$150$\AA\ (rest frame) and up.  Such large equivalent widths are very
hard to achieve with only the ionizing flux from a normal stellar
population, which can provide an equivalent width beyond 200\AA\ for
only a few $\times 10^6$ years (Charlot \& Fall 1993).  To produce equivalent
widths beyond $\sim 200$\AA,
the ultraviolet radiation in the galaxies must be harder
than usual.  This can be achieved in galaxies with active nuclei, with
a stellar metallicity near zero ($Z \ll 0.05 Z_\odot$, Tumlinson \&
Shull 2000; Bromm, Kudritzki \& Loeb 2001), or with a stellar initial
mass function dominated by high mass stars (Malhotra \& Rhoads 2002;
Kudritzki et al. 2000).  However, conventional (broad-lined) active
galactic nuclei are excluded by the observed \lya\ line widths in
these sources, which have upper bounds from $220$ to $460 \kms$
(Table~\ref{spectab}).  Thus, if the sources are
AGN, they must be type~II AGNs, i.e., systems whose
broad-line regions are hidden from direct view.  Even type~II AGN
generally have larger \lya\ line widths than we observe.  For radio
galaxies, a composite spectrum (Stern et al 1999) shows FWHM
$= 1130 \kms$, while an individual case (4c41.17; Dey et al 1997)
shows a narrow component of $613 \kms$ but also a broad component of
similar line flux and FWHM $= 1373 \kms$.  For X-ray selected type~II
AGN, we have one measurement at $1520 \kms$ (Stern et al 2002) and one
upper limit at $< 1130 \kms$ (Norman et al 2002).  Thus, line widths
seem to favor star formation activity over active nuclei as the driver
for the \lya\ line in these LALA sources.

The remaining explanations for the observed equivalent widths both
suggest star formation in primitive conditions.  Low metallicity stars
obviously require low metallicity gas, while top-heavy initial mass
functions may be a natural consequence of the way low-metallicity gas
($Z/Z_\odot \la 5\times 10^{-4}$) cools and fragments (Bromm et al
2001).  A final possibility is that anisotropy in the \lya\ radiation
biases the equivalent width measurements substantially (see Charlot
and Fall 1993; Neufeld 1991), but this explanation would require a
very special source geometry.  Further tests of the narrow-lined AGN
hypothesis are possible, including searches for X-ray emission and for
other optical emission lines characteristic of AGN (especially CIV and
HeII).  Such tests are underway for the LALA survey's larger $z\approx
4.5$ sample.  To date we have seen no obvious evidence for AGN
activity in any of our confirmed high-redshift LALA spectra.

The confirmation of 3 $z\approx 5.7$ \lya\ emitters among 4 candidates
observed allows us to refine the statistics of this population that
were presented in RM01.  There we presented a sample of 18 candidates,
based on detection significance, emission line strength, and an
absence of blue flux.  A complication in candidate selection is that
the measured equivalent widths of high redshift objects may be changed
by intergalactic hydrogen absorption of either continuum or line flux.
A ``worst case'' (and somewhat unrealistic) scenario is where the
broad-band flux is substantially reduced while the narrow-band flux is
unaffected.  If we assume this case and correct the $I$ band fluxes of
our 18 candidates for intergalactic hydrogen absorption using a factor
of $2.1$ (based on the prescription of Madau 1995), we find that 13
candidates still pass the equivalent width cutoff for selection.  The
remaining 5 candidates could in principle be continuum sources at
redshift just below $5.7$, where an apparent narrowband excess of a
factor $\approx 2$ could be produced by IGM hydrogen absorption in the
blue side of the broad band filter.  We will call the 18 candidates
the ``full sample'' and the 13 remaining after our worst case IGM
absorption correction the ``reduced sample'' in the following
discussion.
All four candidates with Keck spectra belong to the reduced sample.

Applying a 3/4 correction factor to the observed source counts for the
spectroscopic success rate presented here, we find a number density of
540 (390) \lya\ emitters per square degree per unit redshift for the
full (reduced) sample. These counts are above a detection threshold of
$(1.47 \pm 0.08) \times 10^{-17} \ergcm2s$, where the error bar
indicates the range in detection threshold between the two narrow-band
filters.

The implied \lya\ luminosity density at $z\approx 5.7$ then becomes
$1.26 \times 10^5$ ($0.80 \times 10^5$) $\lsun/\Mpc^3$ for sources
with $L_\lya > 1.36 \times 10^9 \lsun$, based on the full (reduced)
sample.  (Here and throughout this section we assume a cosmology with
$H_0 = 70 \kmsMpc$, $\Omega_m = 0.3$, and $\Omega_\Lambda = 0.7$.)
This number is a lower bound, and the total luminosity density could
be substantially increased if the luminosity function rises steeply
below our detection threshold.

These \lya\ luminosities can be translated into star formation rates
under an assumed model for the stellar initial mass function (IMF) and
for the escape fraction of \lya\ photons from their parent galaxies.
To enable comparison with other work, we follow Hu, McMahon, \& Cowie
(1999) and apply a conversion factor of $1 \Msun/\year = 10^{42}
\ergsec$.  This factor is obtained by taking the \lya\ to H$\alpha$
ratio for Case B recombination and no dust, together with the Kennicutt
(1983) conversion between H$\alpha$ luminosity and star formation
rate.  This corresponds to an IMF slope $2.5$ (close to the Salpeter
value of $2.35$) over the range $1 < M/M_\odot < 100$, continued to a
lower mass cutoff of $0.1 \Msun$ with slope $1.4$, with solar
metallicity for all stars.  We hasten to add that these assumptions
appear inconsistent with the observed equivalent width distribution of
the LALA \lya\ emitters (Malhotra \& Rhoads 2002), but as we cannot
uniquely determine the correct IMF and metallicity from the available
data, it is not clear how we should replace these assumptions.  The
resulting star formation rate density at $z\approx 5.7$ is $5 (3)
\times 10^{-4} \Msun/\year/\Mpc^3$ in objects with $L_\lya > 7\times
10^{42} \ergsec$, for the full (reduced) samples after correction
for spectroscopic completeness. The star formation rates per object
range from $7$ to $14 \Msun/\year$.

The \lya\ luminosities allow us to constrain the metal production rate
density in these objects much more reliably than the total star
formation rate density.  The ratio of ionizing photon production to
metal production depends only weakly on the details of a stellar
population, because the most massive stars dominate both production
rates (e.g., Madau \& Shull 1996).  Every observed
\lya\ photon implies the production of $\ge 1.5$ ionizing photons
(where equality is expected in the limit of zero dust attenuation).
The metal production rate is approximately $d M_Z / d t = 500 L_i /
c^2$, where $L_i$ is the luminosity in ionizing photons and $M_Z$ is
the mass in elements with atomic number $Z \ge 6$  (Madau \& Shull 1996).
Thus, our \lya\ luminosity density implies a metal production rate
density $\ge 8.6 (5.5) \Msun/\Myr/\Mpc^3$ for the full (reduced) sample.
Malhotra \& Rhoads (2002) suggest that the \lya\
emitters in the LALA sample suffer little dust attenuation, based on
their large observed equivalent widths.  Our lower bound is then
likely to be near the total metal production in \lya\ emitting galaxies.
However, Malhotra \& Rhoads (2002) also note that additional
objects (accounting for $\sim 90\%$ of all galaxies) could
be removed from the sample by dust obscuration.  This would result
in a total metal production rate $\sim 10$ times larger.  For
comparison, we estimate the metal production rate density in
Lyman break galaxies at $z \approx 4$ to be $\sim 14$ times larger
than that in our observed \lya\ sample,
based on the UV luminosity densities in Steidel et al. (1999).
Earlier work by Madau et al (1996), adjusted to our assumed cosmology,
gives $85 \Msun/\Myr/\Mpc^3$ at $z \approx 4$, for continuum selected
450~nm dropouts in the Hubble Deep Field.  This is again a lower bound
that can be increased by corrections for low luminosity sources and
for dust absorption.

Our spectroscopic results also reinforce the conclusion of RM01 that
the space number density of \lya\ emitters with line {\it
luminosity\/} $L_\lya \ga 7 \times 10^{42} \ergsec$ shows little
evolution between $z\approx 4.5$ and $z\approx 5.7$.  In contrast, the
surface density per unit redshift of \lya\ emitters above observed
line {\it flux\/} $\approx 1.5 \times 10^{-17} \ergcm2s$ decreases by
a factor of $\ga 4$ over the same range.  This is in contrast to the
baseline model of Haiman \& Spaans (1999), which would predict a
decrease by at most a factor of $2$ over this redshift range.

Finally, the spectroscopic confirmation of these low-luminosity \lya\
sources at $z\approx 5.7$ secures the conclusion (RM01) that the
reionization redshift $\zre > 5.7$.  The basic physics underlying this
conclusion is simple.  \lya\ photons propagating through intergalactic
gas with a substantial neutral fraction are subject to resonant
scattering by atomic hydrogen.  Thus, even flux on the red side of
line center can be hidden by the red damping wing of a substantially
neutral IGM (Miralda-Escude 1998; Miralda-Escude \& Rees 1998).  This
would obscure the \lya\ line from view, so that detection of the \lya\
line at some redshift $z$ implies a predominantly ionized medium
around the line source, and suggests $\zre > z$ (Haiman \& Spaans
1999).  While the resonantly scattered photons will eventually
redshift in the Hubble flow and escape to infinity, their effective
photosphere (or last scattering surface) is expected to subtend an
angle $\sim 15''$ (Loeb \& Rybicki 1999), reducing their surface
brightness by a factor $\ga 100$ and thereby rendering them
undetectable to surveys like LALA.

This test was first applied by RM01 to show that $\zre > 5.7$, and
subsequently by Hu et al. (2002) to argue for $\zre > 6.56$.  To make
these reionization limits firm, several nuances in the above physical
argument deserve close consideration.  In particular, these objects
have to produce ionizing photons in order to generate the \lya\ line
by recombination.  If the escape fraction $\fesc$ for the ionizing
flux is large, the source will be surrounded by an ionized bubble in
the IGM.  The size of this bubble is limited by the (time-integrated)
ionizing photon production of the source.  RM01 show that the ionized
bubble (in the dust-free case) will be large enough for \lya\ photons
to escape freely (i.e.,
$1.2 \Mpc$) if $L_{43} t_8 \fesc/(1-\fesc) \ga 5$, where
$L_{Ly-\alpha} = 10^{43} L_{43} \ergsec$ and the source is $10^8 t_8$
years old, but that $L_{43} t_8 \fesc/(1-\fesc) \la 2$ for the observed
$z\approx 5.7$ LALA sources.  The critical constraint on $t_8$ and
$\fesc$ is derived from the equivalent widths of the \lya\ sources,
which can only achieve their large observed values in young galaxies
with modest escape fractions.  The Hu et al. (2002) source, HCM~6A, has
a more modest equivalent width ($25$\AA\ rest frame) that could be
produced in systems with $t_8 \sim 1$ and $\fesc \sim 0.75$, but also
shows lower line luminosity $L_{43} \approx 0.3$.  It therefore
fulfills $L_{43} t_8 \fesc/(1-\fesc) \la t_8 \ll 8$, where the last
inequality comes from the age of the universe at this redshift.  This
should be compared with a bound $L_{43} t_8 \fesc/(1-\fesc) \ga 7$ to
produce a Stromgren sphere of radius $1.2\Mpc$ in the denser IGM at
$z=6.56$.  Thus, it again appears unlikely that HCM~6A
could ionize a sufficiently large bubble in a fully neutral IGM to
allow its \lya\ photons to escape freely to infinity.  It is possible,
of course, for any particular \lya\ source near $\zre$ to be more
luminous than we think and have its \lya\ flux substantially
attenuated an effective \lya\ optical depth of a few.  Thus, HCM~6A
al object may be consistent with $\zre \sim 6$ (e.g. Haiman 2002).
However, the \lya\ luminosity function is quite steep at these flux
levels, so the number of detectable \lya\ sources is still expected to
drop sharply at $\zre$.
Haiman (2002) estimates that the suppression of observed \lya\ flux
would be a factor $\sim 10$ if HCM~6a is in a fully neutral IGM.  This
would imply an intrinsic line luminosity $\sim 3\times 10^{43}
\ergsec$, making HCM~6a more luminous than any object found by the
LALA survey in a volume $\sim 30\times$ greater than surveyed by Hu et
al.  We therefore regard this possibility as unlikely.

Other possible loopholes involve the detailed
kinematics of the \lya\ emitters and the neutral IGM, including the
possible effects of gas infall, galactic peculiar motions, and
galactic winds.  Detailed consideration of these effects is beyond the
scope of the current paper.  However, we can make quick estimates of
these effects.  First, infall of the IGM to a newly forming galaxy can
blueshift the gas nearest the galaxy outside the \lya\ line core, but
there must be some point along the line of sight where infalling gas
gives way to the Hubble flow and the IGM has zero velocity relative to
the emitted line photons.  Hence, infall will not greatly reduce the resonant
scattering opacity.  Next, note that the required velocity scale
to allow free escape of \lya\ is $\sim 1000 \kms$.  Peculiar
velocities of this magnitude are restricted to the richest clusters of
galaxies today, and should be vanishingly rare in the $z\sim 6$
universe.  Finally, a spherically symmetric galactic wind with (peculiar)
velocity $\sim 1000 \kms$ at the source, smoothly decreasing to match
the Hubble flow with zero peculiar velocity at $\sim 1.2 \Mpc$, would
require a total energy $\sim 10^{61.5} \erg$.  Spread over a Hubble
time at $z\sim 6$, this corresponds to a mechanical luminosity of
$\sim 10^{11.5} \lsun$, substantially exceeding the total UV/optical light
output of these objects ($\la 10^{11} \lsun$).
The required energy would of course be reduced in an IGM with
substantial clumping or ionization.  However, mechanical energy is
typically $\sim 1\%$ and at most $\approx 10\%$ of the starlight in a
starburst (Leitherer \& Heckman 1995), a factor of $\ga
100$ below the na\"{\i}ve calculation above.  Thus, while galactic
winds are common features of starbursts and probably play a role in
\lya\ escape from local galaxies (Kunth et al 1998), they may not
be able to ``save'' \lya\ emission in a substantially neutral IGM.

A final concern is the possibility that a low-luminosity \lya\ source
may be visible at $z>\zre$ if it is located within the ionized bubble
of a more luminous neighbor.  In this case, the neighbor might be
visible to the \lya\ survey itself or to a Lyman break survey of the
region (such as the NDWFS).  The required luminosity corresponds to
$I_{AB} \la 24-7.5*lg(R/2 \Mpc)$, for a Stromgren sphere radius $R$
and a normal starburst ionizing spectrum.  This number is only a lower
bound because it assumes $\fesc = 1$, which is almost certainly false
for the ionizing photons.  A search for NDWFS sources with $I_{AB} <
24$ and nondetections at the $2\sigma$ level in both $R$ and $B_W$
filters finds no viable candidates within $2 \Mpc$ (projected physical
separation) of any of the three spectroscopically confirmed sources.

In any case, Stromgren spheres of bright neighbors may explain an
isolated \lya\ emitter prior to reionization (such as the Hu et al.
source at $z=6.56$), but these Stromgren spheres
include a minority of the volume before the overlap phase of
reionization.  Thus, a statistical sample of \lya\ emitters such as
the LALA survey provides should still see a drop in counts by a factor
of $\ga 2$ at the overlap phase.  The fact that we see no such drop in
our $z=5.7$ sample (RM01) reinforces the argument
that $\zre > 5.7$ on the line of sight to the LALA Bootes field.

\begin{figure}[ht]
\plotone{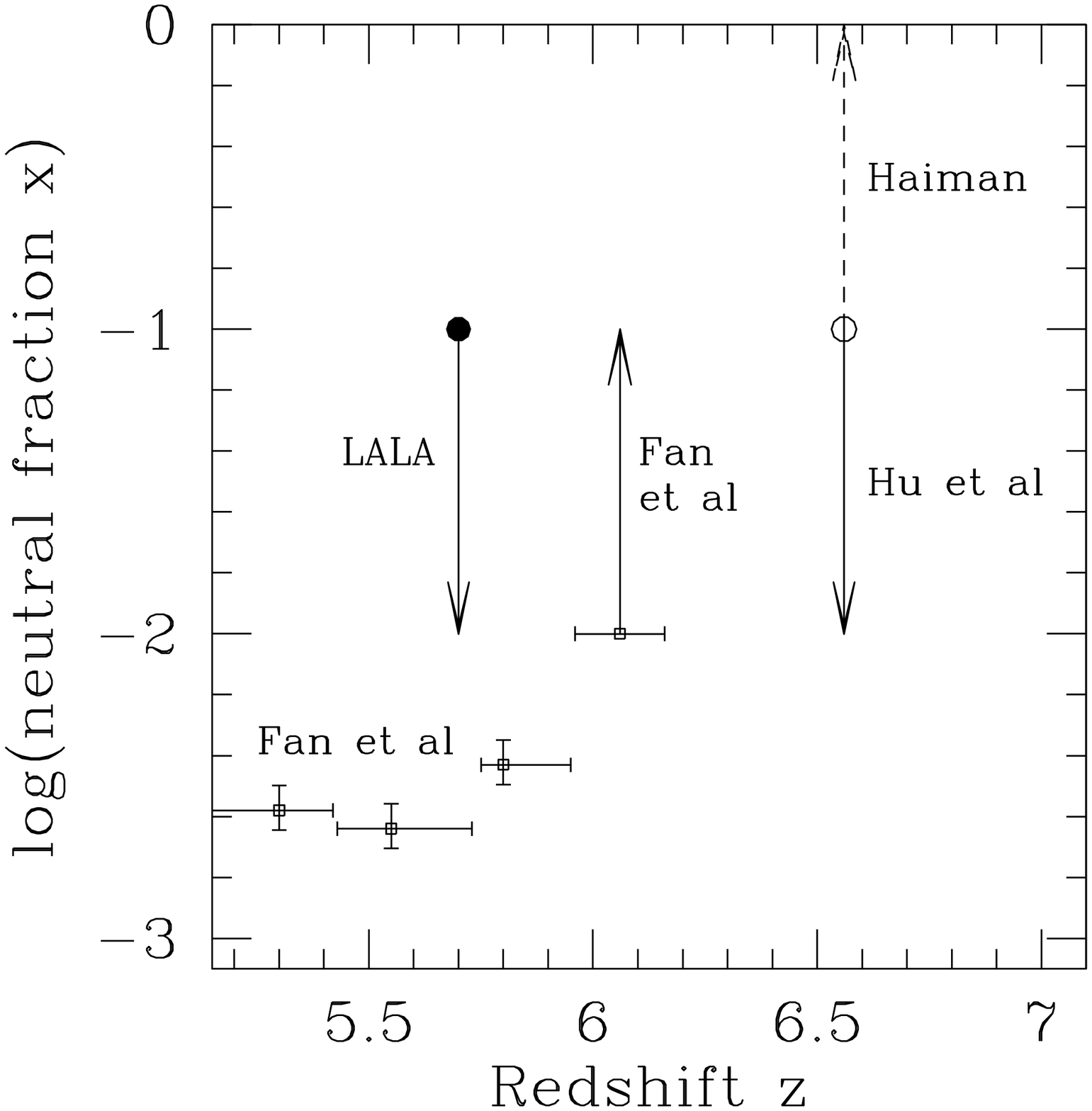}
\caption{
Observational constraints on reionization.  The solid circle shows the
upper limit on the (spatially homogeneous) neutral fraction inferred
from \lya\ galaxy statistics at $z\approx 5.7$ (this paper and
RM01). The open squares with error bars show the evolution of the
mass-averaged neutral fraction $x_{HI}$ that Fan et al (2002) derive
from the \lya\ forest and Gunn-Peterson trough in a sample of quasars
up to redshift $z=6.28$, using an inhomogeneous IGM model.  Finally,
the open circle shows the upper limit to the neutral fraction
suggested by the discovery of a \lya-emitting galaxy at $z=6.56$
(HCM~6a; Hu et al 2002).  Several loopholes might allow this source to be
observed in a predominantly neutral IGM (e.g., Haiman 2002), but we
regard these as unlikely (see section~\ref{disc}).
\label{rei_sum}}
\end{figure}

The neutral fraction required to effectively scatter most of the \lya\
flux from low-luminosity \lya\ sources can be estimated by setting
$\tau \approx 1$ at a velocity $100 \Delta v_{100} \kms$ on the red
side of line center.  RM01 find that this requires a homogeneously
mixed neutral fraction $\approx 0.1 \Delta v_{100}
[(1+z)/6.7]^{-3/2}$.  Figure~\ref{rei_sum} summarizes our present
observational understanding of reionization by combining this estimate
for the neutral fraction limits placed by \lya\ sources with the
mass-averaged neutral fraction evolution that Fan et al (2002) infer
from Gunn-Peterson trough searches in high redshift quasars (Becker
et al 2001; Pentericci et al 2002).  More
detailed modeling that combines the \lya\ luminosity function with
a more realistic, inhomogeneous IGM is planned and will reduce the
uncertainties in figure~\ref{rei_sum}.  Still, because the
Gunn-Peterson trough cannot effectively probe the $x_{HI} \gg 0.01$
regime, it is already clear that the \lya\ emitting galaxies have a role
to play in extending our knowledge of reionization to $z \ga 6.5$.

To summarize, we have presented spectra confirming redshifts $z\approx
5.7$ for three \lya\ emitting galaxies selected from the Large Area
Lyman Alpha (LALA) survey.  These objects were among four candidates
observed from a total sample of 18 candidates.  This implies a total
sample of $\sim 13$ real $z\approx 5.7$ galaxies, easily the largest
uniformly selected galaxy sample at $z > 5.5$.  These \lya\ emitters
all have narrow lines (inconsistent with broad-lined quasars) and very
weak to absent continuum emission.  They belong to the class of high
equivalent width \lya\ sources found in the LALA survey at $z\approx
4.5$.  The population properties of these high equivalent width
objects are reasonably explained by young galaxies with comparatively
low metal and dust content (Malhotra \& Rhoads 2002).  The $z\approx
5.7$ population is similar to the $z\approx 4.5$ one in its number
counts.  This suggests that the sources reside in a predominantly
ionized universe and hence that the reionization redshift $\zre >
5.7$.

\acknowledgements
We thank Gary Puniwai for help with the observations presented here.
JER's research is supported by an STScI Institute Fellowship.
The work of DS was carried out at the Jet Propulsion
Laboratory, California Institute of Technology, under a contract with
NASA.
HS's research was supported by NSF grant 0097163.
The work of SD was was supported by IGPP-LLNL University Collaborative
Research Program grant \#02-AP-015, and was performed under the auspices
of the U.S.\ Department of Energy, National Nuclear Security
Administration by the University of California, Lawrence Livermore
National Laboratory under contract No.\ W-7405-Eng-48.
The data presented herein were obtained at the W.~M. Keck
Observatory, which is operated as a scientific partnership among the
California Institute of Technology, the University of California and
the National Aeronautics and Space Administration.  The Keck Observatory
was made possible by the generous financial support of the W.M. Keck
Foundation.  
The authors wish to recognize and acknowledge the very significant
cultural role and reverence that the summit of Mauna Kea has always had
within the indigenous Hawaiian community.  We are most fortunate to have
the opportunity to conduct observations from this mountain.

\end{document}